\documentclass[12pt]{iopart}

\usepackage{graphicx}
\usepackage{indentfirst}
\usepackage[justification=centering]{caption}
\usepackage[numbers,sort&compress]{natbib}
\usepackage[justification=justified]{caption}
\usepackage{xcolor}
\expandafter\let\csname equation*\endcsname=\relax
\expandafter\let\csname endequation*\endcsname=\relax
\usepackage{amsmath,amssymb,amsthm}
\begin{document}

\title[]{Doppler-induced continuous spectral broadening of ultraviolet lasers}

\author{Huanhuan Wu$^{1,\dagger}$, Yuhan Liu$^{1,\dagger}$, Shengqiang Zhong$^{1}$, Yaozhi Yi$^{1}$, Zhuwen Lin$^{1}$, Hongwei Yin$^{1}$, Yilin Xu$^{1}$, Fan Yang$^{1}$, Xiantao Jiang$^{1,*}$, Yao Zhao$^{1,*}$}

\address{$^1$School of Science, Shenzhen Campus of Sun Yat-sen University, Shenzhen 518107, China}

\address{$\dagger$These authors contributed equally to this work.}
\address{$^*$Author to whom any correspondence should be addressed.}

\ead{jiangxt27@mail.sysu.edu.cn,zhaoyao5@mail.sysu.edu.cn}

\begin{abstract}
We propose a compact scheme based on ultrafast-rotating phase plates (URPPs) to achieve continuous spectral broadening of ultraviolet lasers. The rapid rotation elements behave as a random oscillator which induces Doppler frequency shift into the ultraviolet lasers. As an example, for a disk-shaped phase plate, with the beam acting on the edge at a radius of 10 cm, a rotation frequency of 1 kHz, and a phase-element size of 10 nm, the continuous spectral broadening reaches 0.07\%. Further increasing the rotation speed or reducing the phase-element can lead to greater spectral broadening. When multiple URPPs are arranged in series, the superimposed spatiotemporal modulation further enhances the continuous spectral broadening and achieves more effective speckle smoothing. The scheme is applicable to broadening the independent spectrum of optical frequency combs as well as to the mitigation of laser–plasma instabilities in inertial fusion energy.
\end{abstract}
\maketitle


\indent\setlength{\parindent}{2em} 

Broadband ultraviolet (BUV) light finds important applications in ultrafast spectroscopy, high-resolution microscopy, precision spectroscopic metrology, and inertial fusion energy (IFE) \cite{muraviev2024dual,kirchner2025ultra,hong2023intense,shi2019high,kottig2017generation,song2023ultraviolet,hurricane2023physics}. In ultrafast spectroscopy and high-resolution microscopy, the high spatial uniformity and temporal stability of BUV light help improve measurement accuracy and imaging resolution \cite{furst2024broadband,kotsina2022spectroscopic,mccauley2024dual}. In precision spectroscopic metrology, BUV light enables sub-Hertz frequency resolution and extremely high measurement stability, making it critical for high-precision optical frequency comb calibration and multi-channel synchronous measurements \cite{xu2024near,cingoz2012direct,gohle2005frequency,pupeza2021extreme}. In high-energy-density physical experiments such as IFE, its broadband characteristics can effectively mitigate laser–plasma instability (LPI), improve implosion symmetry, and enhance laser energy coupling efficiency \cite{thomson1974effects,bates2023suppressing,follett2018suppressing,zhao2015effects,lei2024reduction}. However, the proposed schemes for generating BUV still face a series of key bottlenecks. For example, excimer lasers suffer from over 95\% power loss due to the use of masks, which seriously limits its high-throughput applications \cite{jain2005ultrafast,jain1990excimer,obenschain2015high,shen2025design}.

Current mainstream approaches for generation of BUV are based on nonlinear frequency conversion \cite{mutailipu2023achieving,xie2024generation,zhao2025direct,grigutis2023broadband,ning2023broadband,yang2025mechanisms,hickstein2017ultrabroadband,zhang2025octave,kalashnikov2007spatial,darginavivcius2013ultrabroadband}. Low-coherence 529 nm lasers have been generated with conversion efficiency around 63\% \cite{gao2020development,kanstein2025experimental,dorrer2022spectral}. Nevertheless, no practical scheme has been demonstrated up to now for efficient third-harmonic generation (THG) of broadband beams \cite{dorrer2021broadband}. Therefore, existing nonlinear crystal frequency conversion techniques face challenges including low conversion efficiency of THG, limited phase-matching bandwidth, and insufficient optical damage threshold \cite{hosseini2018uv,lekosiotis2020generation,redding2012speckle}. The concept of polychromatic drivers is proposed as a feasible and effective way to generate BUV compared to the conventional broadband driver \cite{zhao2022polychromatic,zhao2023control}. High-power ultraviolet light can easily induce surface or bulk damage in crystals, posing severe challenges for large-aperture. High-uniformity crystals are not only costly to fabricate but also have limited energy-handling capabilities. To improve the conversion rates and bandwidth limitations of conventional crystal-based approaches, plasma-based modulation schemes were proposed \cite{yu2016plasma,zhao2020plasma,geng2018enhancement,ganeev2014quasi,kuo2007enhancement,rodel2012harmonic,mathijssen2023material}. However, technical challenges and insufficient stability continue to hinder its widespread application, while also suffering from large beam divergence and unstable beam quality.

Various beam-smoothing techniques have been proposed to improve beam uniformity and reduce local peak intensity. For example, continuous phase plates (CPPs) spatially smooth the Gaussian beams through generation of static speckle \cite{yang2008continuous,yang2013novel}. Spectral dispersion smoothing (SSD) exploits the coupling between frequency modulation and angular dispersion, causing different frequency sub-beams to generate rapidly evolving interference patterns on the focal plane, which forms highly uniform focal spots after temporal averaging \cite{skupsky1989improved,fan2025numerical,dorrer2025demonstration,fusaro2024improvement}. Although these techniques have made important progress in beam smoothing, efficient generation of BUV light in compact optical systems remains challenging.

Here, we propose a Doppler-induced continuous spectral broadening scheme based on URPPs, which introduces time-varying random phases to simultaneously smoothing the spatial structure of the laser beams. Theoretical and simulation results demonstrate that URPPs can effectively enhance both spectral broadening and beam uniformity. The smoothing effects can be enhanced by increasing the rotation speed, decreasing the size of random phase elements, and employing cascaded URPPs. This scheme features a compact configuration and strong compatibility, allowing it to be integrated as an independent module into the high-power facilities without modifying core systems. The entire broadening process relies on phase modulation with cost-effective and high efficiency, thereby reducing the need for nonlinear crystals. Moreover, the output bandwidth can be flexibly tuned by adjusting the rotation speed and phase plate elements.


According to statistical optics theory \cite{goodman2015statistical}, the autocorrelation function of an optical field is defined as
\begin{align} 
C(\tau') &= \left\langle E(\mathbf{x},\mathbf{y}, t) \, E^*(\mathbf{x},\mathbf{y}, t+\tau') \right\rangle_t \nonumber\\
&= E^2 e^{i \omega_0 \tau'} \, 
\left\langle e^{\,i[\phi(\mathbf{x},\mathbf{y},t)-\phi(\mathbf{x},\mathbf{y},t+\tau')]} \right\rangle_t. \label{eq:autocorrelation}
\end{align} 
where $E = E_0 \exp[-2(x^2+y^2)/W]$ is the amplitude of the incident laser beam, $\omega_0$ is the central frequency, and $W$ is the waist radius. $\phi(\mathbf{x},\mathbf{y},t)$ denotes the instantaneous optical phase, and the notation $\langle \cdot \rangle_t$ indicates temporal averaging. Taking the Fourier transform of the autocorrelation function yields the power spectral density of the output optical field:
\begin{align}
|F(\omega)|^2 &= \mathcal{F}\{C(\tau')\} 
= E^2 \, \mathcal{F}\{ e^{i \omega_0 \tau'} \} * 
\mathcal{F} \!\left\{ \mathrm{tri}\!\left(\frac{\Omega r \tau'}{d_r}\right) \right\} \nonumber,\\
\end{align}

where $\mathcal{F}{\cdot}$ denotes the Fourier transform.
About 76.1\% of laser energy is contained within the full width at half maximum of a Gaussian spectrum, therefore we define the effective spectral width based on this energy fraction. According to Eq.~(2), the effective bandwidth of $|F(\omega)|^2$ is
\begin{align}
\Delta \omega = \frac{6.08 \, r \, \Omega}{d_r},
\end{align}
Based on eq. 3, for a disk-shaped phase plate, with the beam acting on the edge at a radius of 10 cm, a rotation frequency of 1 kHz, and a phase-element size of 10 nm, the continuous spectral broadening reaches 0.07\%. By modulating the laser wavefront with a single URPP (effectively a moving optical element), fast phase modulation is introduced into the temporal domain, which is equivalent to generating a Doppler shift. Random patterns within the beam can scatter the single-frequency laser into a continuous spectrum with a finite bandwidth. The output spectral width is proportional to the angular velocity $\Omega$, and the radial position of the beam $r$, while being inversely proportional to the size of phase elements $d_r$. The modulation frequency of a single URPP is $r\,\Omega/d_r$ at $r$. We adopt an area-integration method to obtain the average modulation frequency of the overall laser spot. Consequently, the modulation frequency as a function of the phase plate rotation speed and beam radius is given by ${f_m} = 2\Omega R/3d_r,$ where $R$ is the effective beam radius, ${f_m/\omega_0}$ is the normalized modulation frequency. This indicates that increasing the rotation speed, enlarging the beam radius, or reducing the size of the phase elements can all enhance spectral broadening.


\hfill
\begin{figure}[htb!]
\vspace{0cm}
\setlength{\abovecaptionskip}{-0.8cm}
\begin{center}
\centerline{\includegraphics[height=8cm,width=15cm]{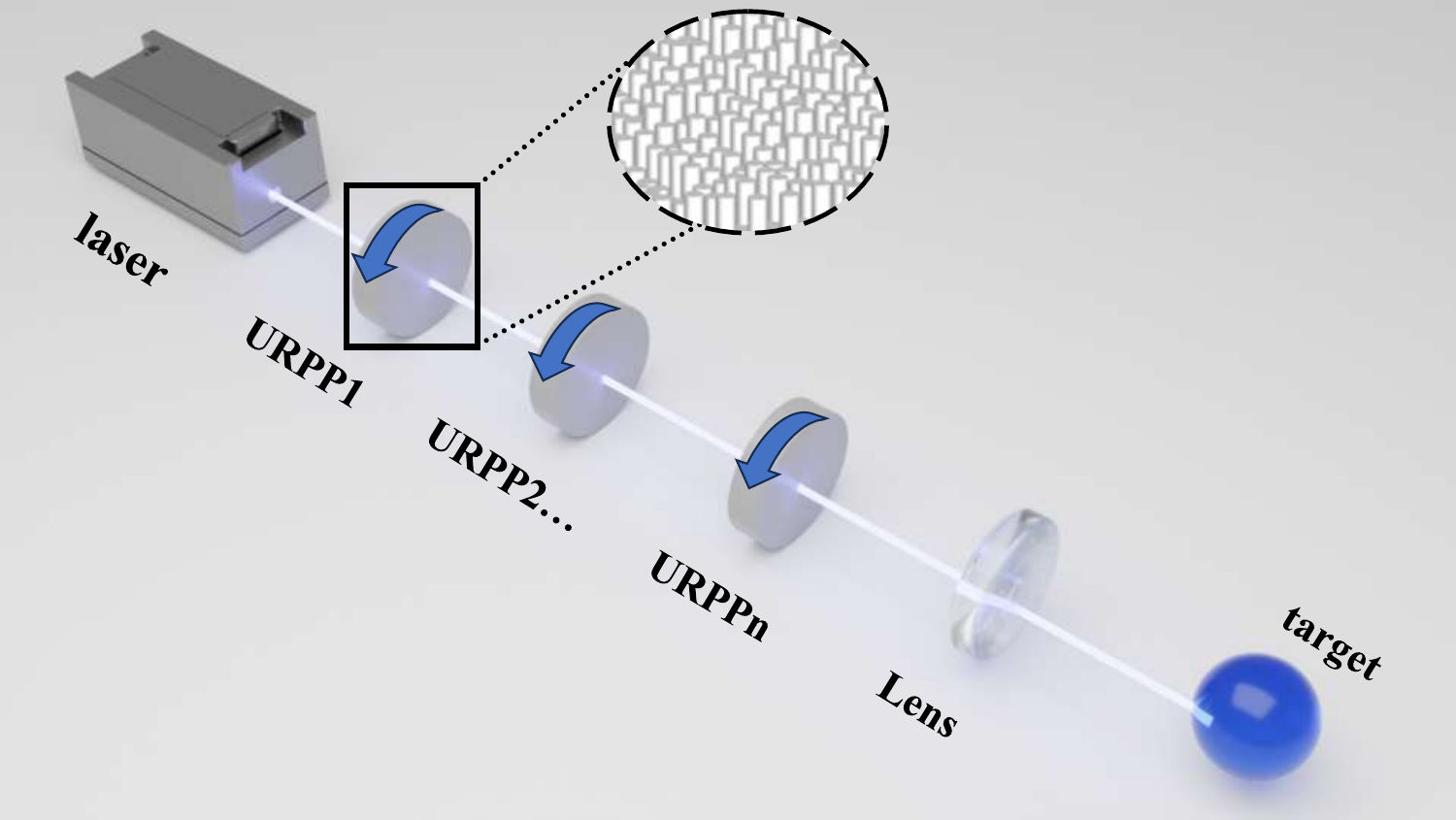}}
\end{center}
\caption{(color online) Schematic of cascaded spectral broadening using multiple URPPs, where the laser sequentially passes through each URPP with a random height structure.}
\vspace{0.3cm}
\label{fig:fig1}
\end{figure}

Figure~1 shows the optical layout consisting of multiple URPPs for a cascade spectrum broadening. A Gaussian laser pulse sequentially passes through each URPP, where spatiotemporal phase modulations are imposed on the beam, thereby perturbing its wavefront. The modulated beam is then focused by a lens, forming a dynamic speckle pattern as well as spectral broadening. A static URPP behaviors like a CPP.

\hfill
\begin{figure}[htbp]
\vspace{-1cm}
\setlength{\abovecaptionskip}{-0.8cm}
\begin{center}
\centerline{\includegraphics[height=13cm,width=17cm]{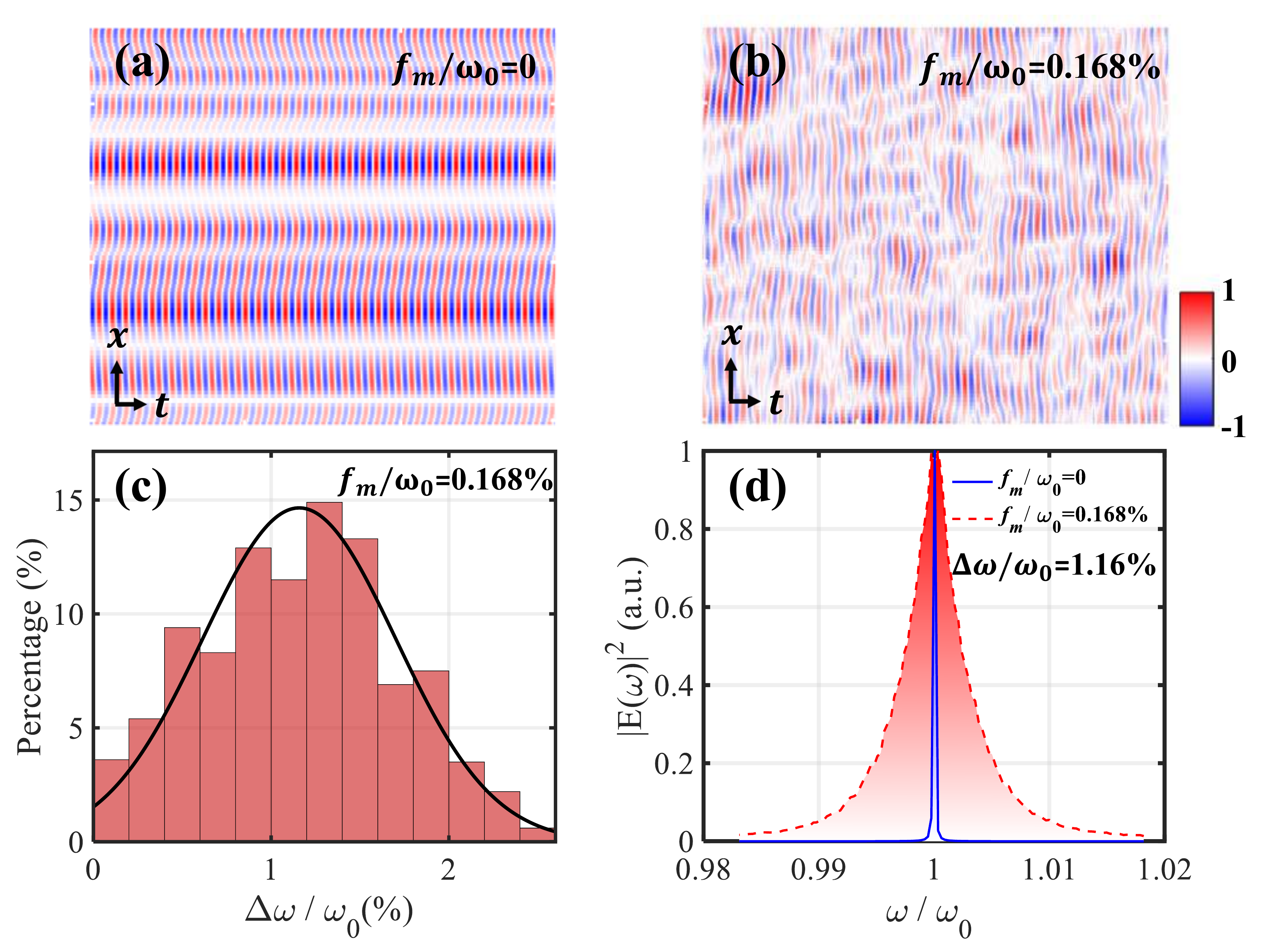}}
\end{center}
\caption{(color online) Laser speckle distributions for (a) $f_m/\omega_0 = 0$, and (b) $f_m/\omega_0 = 0.168\%$, where the horizontal axis represents time $t$ ($fs$) and the vertical axis represents transvers axis $y/\lambda_0$. (c) Statistical histogram of spectral broadening $\Delta\omega/\omega_0$. (d) Comparison of the spectra for $f_m/\omega_0 = 0$ and $f_m/\omega_0 = 0.168\%$.}
\vspace{0.3cm}
\label{fig:fig2}
\end{figure}

The optical field after passing through a single URPP is expressed as  $E_{\rm out}(x,y,t)=E \, e^{-i \omega t} \, e^{i \phi(x,y,t)}$, and the subsequent focusing process with a lens can be described using Fresnel diffraction:
\begin{equation}
E_{\rm focal}(x',y',t) = 
\frac{e^{i k f}}{i \lambda f} \,
\exp\!\Big[\frac{i k}{2 f} (x'^2 + y'^2)\Big] \,
\mathcal{F}_{x,y}\!\{ E_{\rm out}(x,y,t) \},
\end{equation}
where $E_{\rm focal}(x',y',t)$ represents the final speckle field of the focused broadband laser. 
A temporal Fourier transform of $E_{\rm focal}(x',y',t)$ yields the intensity spectrum. The numerical results are shown in Fig. 2, where the time window is $4~\mathrm{ps}$, the numerical results of a UV laser with wavelength 351 nm are shown, the phase-element size is $d_r = 5.4~\mathrm{\mu m}$, the effective beam radius is $R = 1~\mathrm{cm}$, the output spot size is 650 um, and the temporal and spatial resolutions are $\Delta t = 0.08~\mathrm{fs}$ and $\Delta x = \Delta y = 0.0325~\mathrm{\mu m}$, respectively.

Figures~2(a) and 2(b) show the time-varying electric fields for $f_m/\omega_0 = 0$ and $f_m/\omega_0 = 0.168\%$, respectively. Static speckle fields can be found for $f_m/\omega_0 = 0$ owing to the narrow spectrum, whereas the speckle distribution changes randomly and rapidly over time at $f_m/\omega_0 = 0.168\%$ due to the spectrum broadening \cite{goodman2007speckle}. Figure~2(c) presents the sampled spectral broadening $\Delta\omega/\omega_0$ at randomly selected positions on the focal plane. The local spectral width $\Delta\omega$ varies across the beam, we obtain the average spectral width of the entire focal spot by sampling different positions and statistically analyzing the distribution of these local spectral widths, with the spectral-width distribution being approximately Gaussian. Figure~2(d) shows the spectral distributions of smoothed lasers with $f_m/\omega_0 = 0$ and $f_m/\omega_0 = 0.168\%$. Compared with $f_m/\omega_0 = 0$, a spectral broadening of 1.16\% is observed at $f_m/\omega_0 = 0.168\%$, showing that dynamic phase modulation can effectively broaden the spectrum of the optical field.

\begin{table}
\caption{Spectral broadening under different modulation frequencies with a single URPP}
\centering
\begin{tabular*}{0.6\textwidth}{l @{\extracolsep{\fill}} c} 
\hline
$f_m/\omega_0$ (\%) & $\Delta\omega/\omega_0$ (\%) \\
\hline
0.019 & 0.16  \\
0.037 & 0.29  \\
0.056 & 0.41  \\
0.084 & 0.60  \\
0.112 & 0.78  \\
0.168 & 1.16  \\
\hline
\end{tabular*}
\label{tab1}
\end{table}

\hfill
\begin{figure}[htbp]
\vspace{-0.2cm}
\setlength{\abovecaptionskip}{-0.8cm}
\begin{center}
\centerline{\includegraphics[height=7cm,width=17cm]{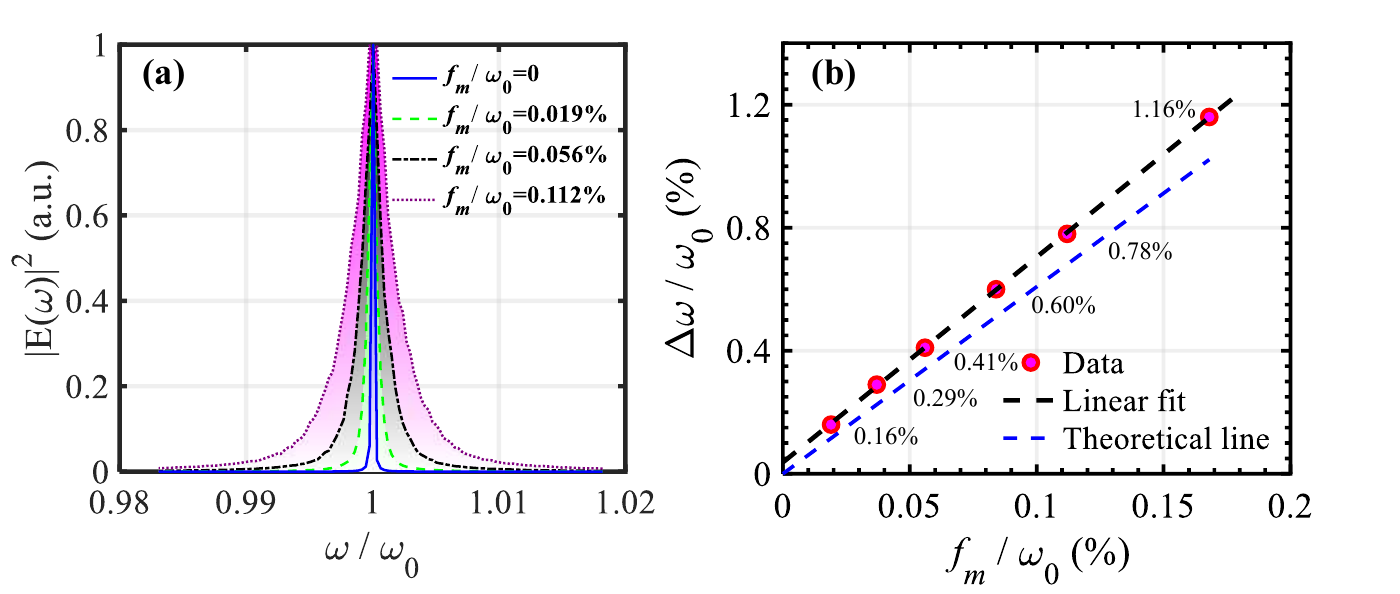}}
\end{center}
\caption{(color online) (a) Comparison of spectra with different modulation frequencies, where the blue, green, black, and purple curves correspond to $f_m/\omega_0 = 0, 0.019\%$, $0.037\%$, and $0.056\%$, respectively. (b) Comparison between theoretical and numerical results of broadening bandwidth varying with modulation frequency, where the focal spot size is 650 um.}
\vspace{0.3cm}
\label{fig:fig3}
\end{figure}

Figure~3(a) shows the spectral distributions at different modulation frequencies, where the blue curve corresponds to the narrowband spectrum for $f_m/\omega_0 = 0$, while the green, black, and purple curves correspond to $f_m/\omega_0 = 0.019\%$, $f_m/\omega_0 = 0.056\%$, and $f_m/\omega_0 = 0.112\%$, respectively. It is demonstrated that faster dynamic modulation effectively produces a broader bandwidth. Figure~3(b) displays the bandwidth of a smoothed laser as a function of the modulation frequency, where the blue dashed line is theoretical results of the smoothed field after passing through a single URPP according to Eq. (3). The red dots demonstrate simulation results of the optical field shaped by a single URPP and subsequently focused by a lens, which fits well with the black linear line. Actually, the lens used in numerical simulations only modulates the spatial phase of laser field, which cannot introduce additional frequency modulations. However, the overall bandwidth is relatively larger than the theoretical results due to the focused energy spectrum arising from spatial interference. Correspondingly, Table~1 presents the exact data of broadening bandwidths caused by the modulation effect of a single URPP. The effective bandwidth $\Delta \omega/\omega_0$ enhances from 0.16\% to 1.16\%, as the modulation frequency increases from $f_m/\omega_0 = 0.019\%$ to $f_m/\omega_0 = 0.168\%$. This indicates that the modulation frequency is a major determinant for controlling the spectral width of the output laser, and higher-frequency phase modulations can effectively reduce the temporal coherence of the optical field.

\begin{table}
\caption{Spectral broadening for cascaded URPPs with different modulation frequency combinations.}
\centering
\begin{tabular*}{\textwidth}{l @{\extracolsep{\fill}} c} 
\hline
$f_m/\omega_0$ (\%) & $\Delta\omega/\omega_0$ (\%) \\
\hline
$f_{m1}=$0.056 and $f_{m2}=$0.037        & 0.66  \\
$f_{m1}=$0.056 and $f_{m3}=$0.084        & 0.97  \\
$f_{m1}=$0.056 and $f_{m4}=$0.112       & 1.18  \\
$f_{m1}=$0.056 and $f_{m5}=$0.168       & 1.30  \\
$f_{m0}=$0.019, $f_{m1}=$0.056, $f_{m3}=$0.084, and $f_{m4}=$0.112       & 1.75  \\
\hline
\end{tabular*}
\label{tab2}
\end{table}

Table~2 shows the cascaded spectral broadening of a laser beam after passing through multiple URPPs with different modulation frequency combinations. Compared with Table~2, it is evident that using two URPPs will result in an increase of spectral bandwidth. For example, when the modulation frequencies of two URPPs are $f_{m1}/\omega_0 = 0.056\%$ and $f_{m2}/\omega_0 = 0.037\%$, the effective bandwidth is 0.66\%, which is larger than that of the single URPP with $f_{m}/\omega_0 = 0.056\%$. With increasing the modulation frequency of the second URPP can further enhance the bandwidth of smoothed lasers. Note that the combination of four URPPs with $f_{m0}=0.019\%$, $f_{m1}=0.056\%$, $f_{m3}=0.084\%$, and $f_{m4}=0.112\%$ results in a relatively larger bandwidth of 1.75\%, indicating a significantly enhanced broadening effect. These results demonstrate that both increasing the modulation frequency of a single URPP or employing multiple URPPs with different modulation frequencies can effectively improve the efficiency of spectral broadening.

\hfill
\begin{figure}[htbp]
\vspace{-1.5cm}
\setlength{\abovecaptionskip}{-1.2cm}
\begin{center}
\centerline{\includegraphics[height=13cm,width=17cm]{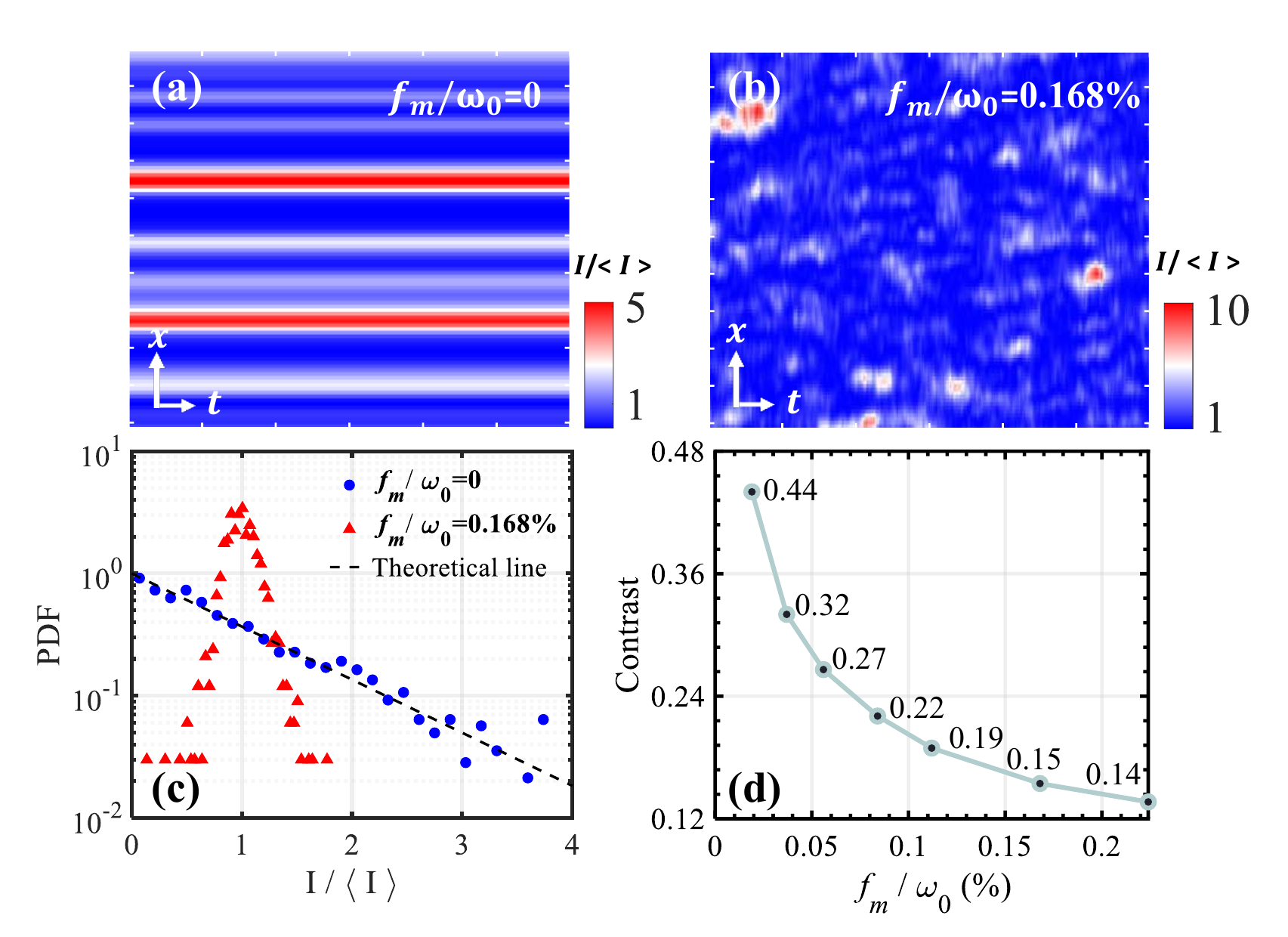}}
\end{center}
\caption{(color online) Spatiotemporal speckle distributions for (a) $f_m/\omega_0 = 0$, and (b) $f_m/\omega_0 = 0.168\%$, respectively. (c) Probability density function (PDF) of the normalized intensity for different modulation frequencies. The blue and red curves correspond to $f_m/\omega_0 = 0$ and $f_m/\omega_0 = 0.168\%$, respectively, while the black line shows the theoretical distribution of Rayleigh speckle. (d) Spatial uniformity of laser speckle smoothed by a single URPP with different modulation frequencies.}
\vspace{0.3cm}
\label{fig:fig4}
\end{figure}

As illustrated in Fig.~4(a), the spatiotemporal intensity distribution for $f_m/\omega_0 = 0$ exhibits regular and stable interference fringes, indicating a high degree of temporal coherence. Figure~4(b) exhibits the speckle distribution for $f_m/\omega_0 = 0.168\%$, where the speckle evolves rapidly over time, resulting in a spatiotemporal random distribution of intensity. To investigate the spatial uniformity of the light field, we first integrate the intensity at each spatial position over time, and then calculate the probability density function (PDF) of the spatial intensity distribution. Figure~4(c) presents the normalized intensity of PDF for different modulation frequencies. The theoretical line shows a negative exponential distribution of Rayleigh speckle, which fits well with the PDF of $f_m/\omega_0 = 0$, exhibiting significant spatial fluctuations \cite{michel2023introduction}. In contrast, the PDF of $f_m/\omega_0 = 0.168\%$ concentrates around $I/\langle I\rangle \approx 1$, indicating effective mitigation of spatial intensity fluctuations, where  $\langle I\rangle$ is an average intensity. To quantify the spatial uniformity, we conduct statistical measurements of the speckle contrast based on $\sigma_I/\langle I\rangle$, where a lower contrast corresponds to a more uniform light field \cite{liu2021generation}. As shown in Fig.~4(d), increasing $f_m$ can significantly reduce the contrast and improve the spatial uniformity of the light field.

\hfill
\begin{figure}[htbp]
\vspace{-1cm}
\setlength{\abovecaptionskip}{-0.8cm}
\begin{center}
\centerline{\includegraphics[height=13cm,width=16cm]{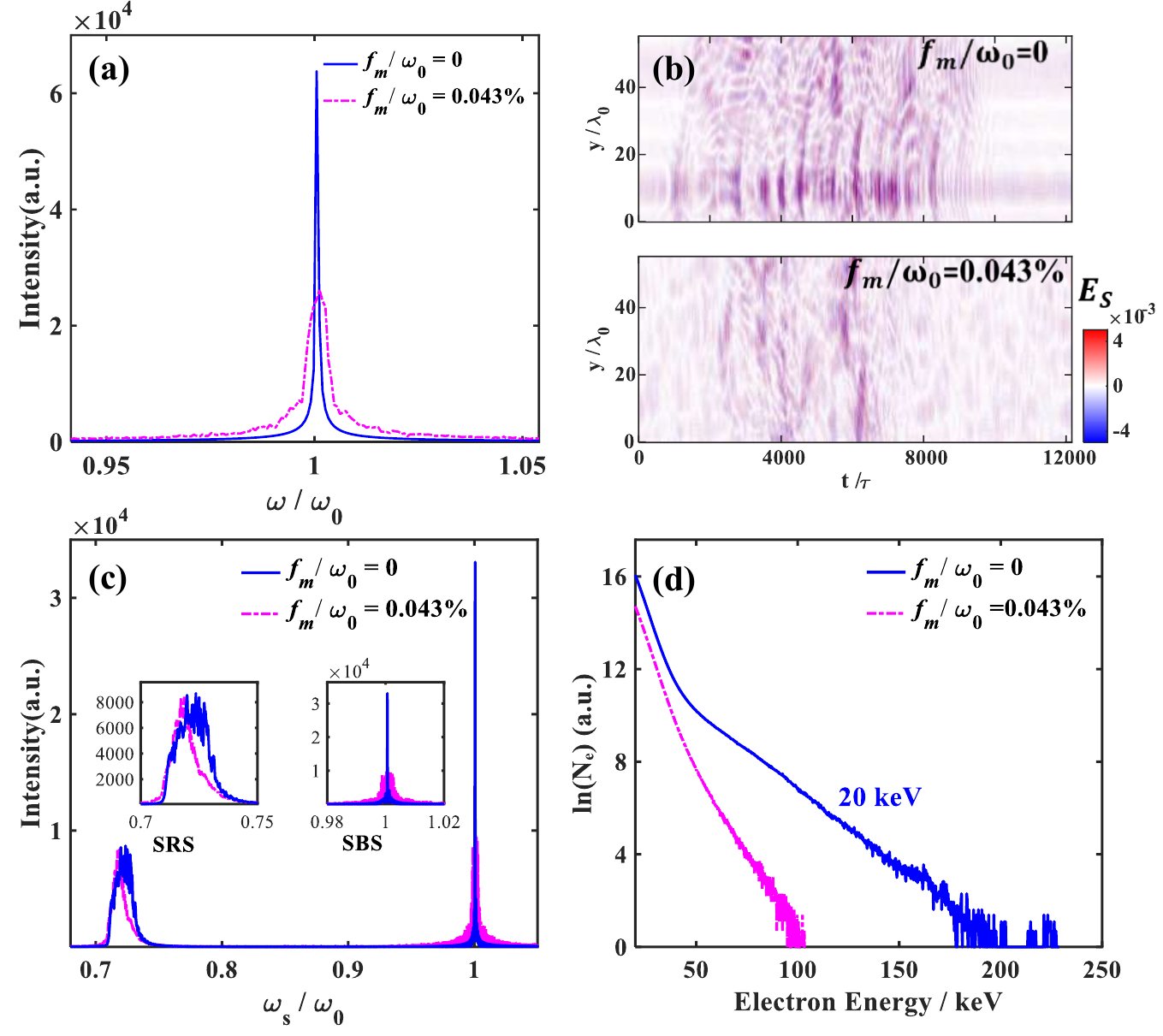}}
\end{center}
\caption{(color online) (a) Spectra of the incident laser, where the blue and red curves correspond to lasers smoothed by a URPP with \( f_m/\omega_0 = 0 \) and \( f_m/\omega_0 = 0.043\% \), respectively. (b) Spatiotemporal speckle distributions for the cases of \( f_m/\omega_0 = 0 \) and \( f_m/\omega_0 = 0.043\%\), where \(\tau\) is the laser period. (c) Spectra of the scattered lights, with colors consistent with panel (a). (d) Comparison of the electron energy spectra heated by two lasers smoothed by a URPP with different modulation frequencies.}
\vspace{0.3cm}
\label{fig:fig6}
\end{figure}

As an application example, the effectiveness of spectral broadening in mitigating laser plasma instabilities is investigated via two-dimensional particle-in-cell (PIC) simulations using EPOCH \cite{arber2015contemporary}. The incident laser is s-polarized, with an average intensity of $1.13\times10^{15}~\mathrm{W/cm^2}$ and wavelength $\lambda_0 = 0.351~\mu\mathrm{m}$. The plasma density is homogeneous with $n_e = 0.08 n_c$, where $n_c$ is the critical density. The longitudinal length of simulation box is $110~\mu\mathrm{m}$, with $5~\mu\mathrm{m}$ left and right vacuum, respectively. Each cell contains 350 macro-particles, and the spatial resolutions are $dx/\lambda_0 = 0.071$ and $dy/\lambda_0 = 0.28$, respectively. The initial electron and ion temperatures are $T_e = 1~\mathrm{keV}$ and $T_i = 0.3~\mathrm{keV}$, respectively. Figure~5(a) shows the spectral distributions of incident lasers smoothed by a URPP with $f_m/\omega_0 = 0$ and $f_m/\omega_0 = 0.043\%$. The distributions of backscattered lights are displayed in Fig.~5(b). The monochromatic laser with static speckle drives intense stimulated Raman (SRS) and Brillouin scattering (SBS), whereas the scattered light is relatively uniform in the transverse distribution for the broadband case with $f_m/\omega_0 = 0.043\%$ \cite{zhong2020improvement}. Meanwhile, a noticeable delay in the excitation of backscattered light can be observed for the broadband case. Figure~5(c) shows the spatially integrated backscattered spectra. The intensity of SBS is significantly reduced for lasers smoothed with $f_m/\omega_0 = 0.043\%$, where the scattering spectrum is around the incident frequency, i.e., $\omega_s / \omega_0 \sim 1$. The SRS intensity is also reduced, which leads to the mitigation of hot-electron generation as shown in Fig.~5(d). A hot tail with electron temperature $\sim$20 keV can be found for the case of monochromatic laser. However, little electrons are heated by the smoothed laser with $f_m/\omega_0 = 0.043\%$ according to the red line. Therefore, LPI can be effectively mitigated by the lasers smoothed by URPPs with high modulation frequencies.


In this work, we propose a scheme to broaden the laser spectrum via Doppler effect induced by random oscillations of phase elements on the URPPs, achieving also the beam smoothing of ultraviolet light. The results show that increasing the modulation frequency of a single URPP or cascading multiple URPPs can further enhance spectral broadening, while significantly improving beam spatial uniformity. The scheme can be integrated into high-power laser facilities without modifying the core optical system, thus providing high energy efficiency and good system compatibility. PIC simulations demonstrate that URPP-smoothed laser beams with high modulation frequency can effectively mitigate laser plasma instabilities, offering a feasible approach for employing broadband ultraviolet light in IFE.

\section*{Acknowledgments}
This work is supported by the National Natural Science Foundation of China (No. 12005287) and the Guangdong Provincial Natural Science Foundation (Grant No. 2024A1515011945). The EPOCH code used in this work was in part funded by the UK EPSRC Grants EP/G054950/1, EP/G056803/1, EP/G055165/1, EP/M022463/1, and EP/P02212X/1.

\section*{Data availability statement}
All data that support the findings of this study are included within the article (and any supplementary files).

\bibliographystyle{iopart-num}
\bibliography{main}

\end{document}